\documentclass[sigconf]{acmart}
\usepackage{tabularx}
\AtBeginDocument{%
  }

\copyrightyear{2026}
\acmYear{2026}
\setcopyright{cc}
\setcctype{by}
\acmConference[WSDM '26]{Proceedings of the Nineteenth ACM International Conference on Web Search and Data Mining}{February 22--26, 2026}{Boise, ID, USA}
\acmBooktitle{Proceedings of the Nineteenth ACM International Conference on Web Search and Data Mining (WSDM '26), February 22--26, 2026, Boise, ID, USA}
\acmPrice{}
\acmDOI{10.1145/3773966.3779388}
\acmISBN{979-8-4007-2292-9/2026/02}

\begin{document}

\title{Cold-Starting Podcast Ads and Promotions with Multi-Task Learning on Spotify}


\author{Shivam Verma}
\authornote{Both authors contributed equally to this research.}
\affiliation{%
  \institution{Spotify}
  \city{London}
  \country{UK}
}

\author{Hannes Karlbom}
\authornotemark[1]
\affiliation{%
  \institution{Spotify}
  \city{Stockholm}
  \country{Sweden}
}

\author{Yu Zhao}
\affiliation{%
  \institution{Spotify}
  \city{Stockholm}
  \country{Sweden}
}

\author{Nick Topping}
\affiliation{%
  \institution{Spotify}
  \city{Seattle}
  \country{USA}
}

\author{Vivian Chen}
\affiliation{%
  \institution{Spotify}
  \city{San Francisco}
  \country{USA}
}

\author{Kieran Stanley}
\affiliation{%
  \institution{Spotify}
  \city{Paris}
  \country{France}
}

\author{Bharath Rengarajan}
\affiliation{%
  \institution{Spotify}
  \city{San Francisco}
  \country{USA}
}

\renewcommand{\shortauthors}{Shivam Verma et al.}

\begin{abstract}

We present a unified multi-objective model for targeting both advertisements and promotions within the Spotify podcast ecosystem. Our approach addresses key challenges in personalization and cold-start initialization, particularly for new advertising objectives. By leveraging transfer learning from large-scale ad and content interactions within a multi-task learning (MTL) framework, a single joint model can be fine-tuned or directly applied to new or low-data targeting tasks, including in-app promotions. This multi-objective design jointly optimizes podcast outcomes such as streams, clicks, and follows for both ads and promotions using a shared representation over user, content, context, and creative features, effectively supporting diverse business goals while improving user experience. 

Online A/B tests show up to a $22\%$ reduction in effective Cost-Per-Stream (eCPS), particularly for less-streamed podcasts, and an $18$--$24\%$ increase in podcast stream rates. Offline experiments and ablations highlight the contribution of ancillary objectives and feature groups to cold-start performance. Our experience shows that a unified modeling strategy improves maintainability, cold-start performance, and coverage, while breaking down historically siloed targeting pipelines. We discuss practical trade-offs of such joint models in a real-world advertising system.

\end{abstract}



\maketitle

\section{Motivation}

\begin{figure}[h]
\begin{tabular}{ll}
\includegraphics[scale=0.35]{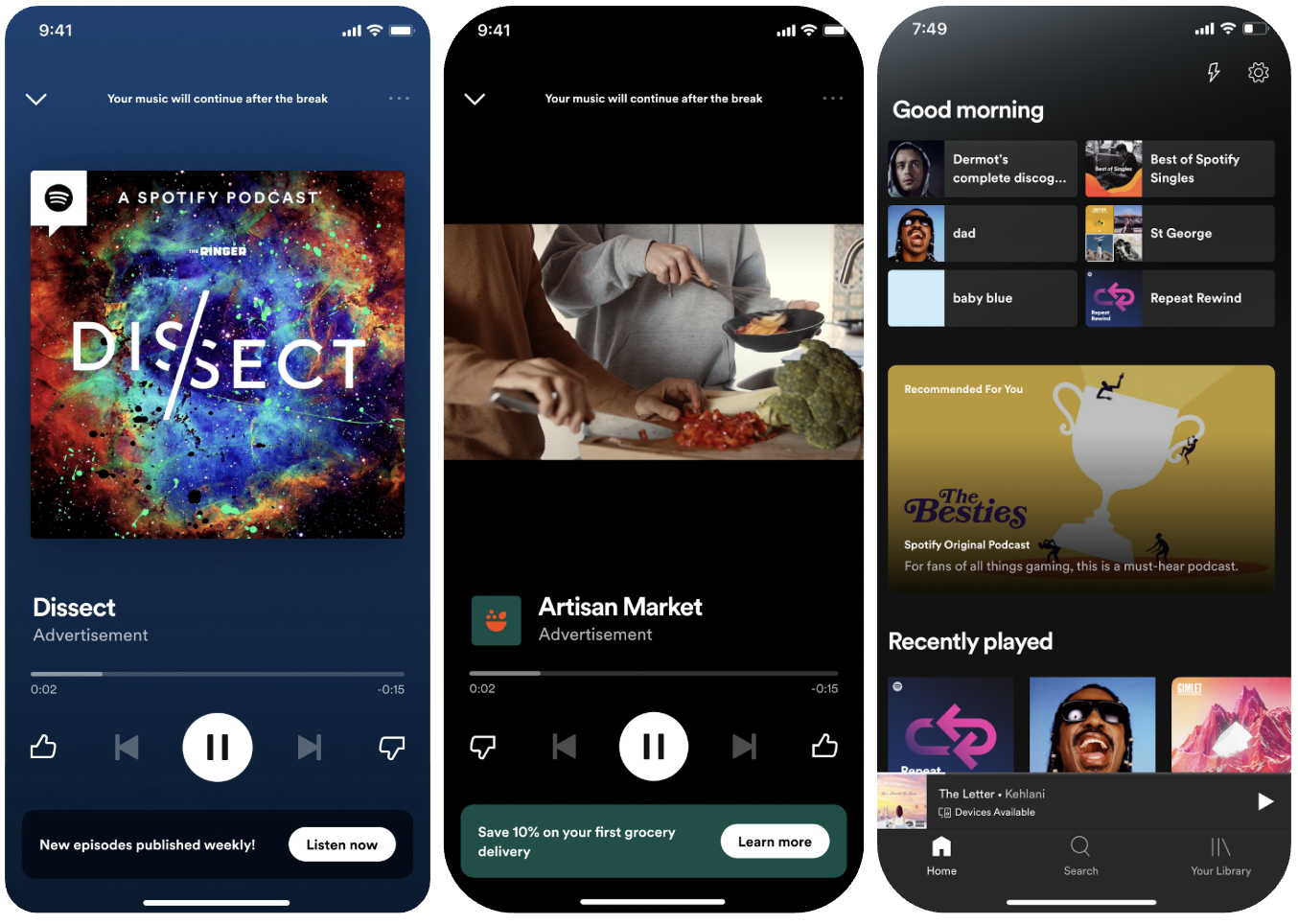}
\end{tabular}
\caption{Left to right: (a) An in-stream podcast audio ad. (b) An in-stream unmuted podcast video ad. (c) A display promotion for a Spotify Original podcast.}
\label{fig:fig1}
\end{figure}

Spotify, with its user base of over 700 million, identifies podcasts as a significant and rapidly growing content vertical, making effective personalization crucial for listener engagement, monetization (especially for over 400 million ad-supported users), and the discovery and growth of podcast creators. The platform supports diverse business objectives, from driving initial streams for new episodes to optimizing impression-to-stream rates (i2s) and click-through rates (CTR), with a particular focus on boosting visibility for less-streamed creators who suffer from cold-start issues due to data scarcity. Two primary mechanisms connect users with podcast content: \textbf{advertisements (ads)}, such as in-stream audio or video placements (Fig.~1a, 1b), and \textbf{promotions}, which surface strategically important or relevant content like display promotions for Spotify Originals (Fig.~1c). Despite different immediate objectives (an ad click versus a direct stream from a promotion), both channels share the goal of matching users with relevant and engaging podcasts, driven by similar user signals (e.g., listening history, explicit follows) and content affinities (e.g., genre, topics). Positive interactions in one channel can therefore inform decisions in the other.

Historically, these objectives were handled by separate, specialized machine learning models. For example, a model optimizing \textit{i2s} for a Home-page promotion would be distinct from a model optimizing clicks on an audio ad for a new podcast series. This siloed, task-specific approach created several challenges. First, \textbf{slow innovation}: introducing new business or ad objectives requires building new models from scratch, involving substantial engineering, data collection, and A/B testing, which can slow the rollout of tools that help podcasters reach relevant audiences. Second, the \textbf{cold-start problem for optimization objectives}: newly introduced ad or promotional products for specific audiences, such as the ``likelihood to stream advertised content after an ad'' for emerging or new creators, often lack sufficient interaction data to train high-performing specialized models, hampering the discoverability of these less-streamed creators. Third, \textbf{inefficiency and missed synergies}: separate pipelines made it difficult to exploit shared latent patterns and overlapping signals across podcast ads and promotions, and led to siloed teams building similar models for related products.

These challenges motivated our exploration of a \textbf{unified objective optimization approach} via multi-task learning (MTL).

\section{Related Work}

Multi-task learning improves performance by jointly learning related tasks \cite{Covington2016, Collobert2008, Ma2018, Long2017, Lin2019, Sener2018, cheng2016wide, ma2018esmm, yi2019sampling}, facilitating transfer from data-rich advertising tasks to data-scarce promotional ones (and vice versa), thereby addressing cold-start issues for newer and smaller creators. By modeling ads and promotions together, we aim to consolidate learning, reduce duplication across systems, and accelerate new capability deployment, ultimately improving personalization for listeners and growth for creators. Our contribution focuses on bridging organizational silos by grouping tasks based on business goal alignment in multi-stakeholder and multi-objective settings \cite{Sener2018, Standley2020, googleads2022, moo1, moo2, zheng2022multiobjective, jannach2023survey}, which is crucial for balancing diverse business objectives.

\textbf{Multi-Task Learning in Industry Recommenders.}
MTL improves generalization by jointly learning related objectives \cite{Caruana1997}. At industrial scale, platforms have adopted MTL to couple heterogeneous business goals such as engagement, satisfaction, and monetization \cite{airbnb2023, lirank2024, verma2025, Covington2016, Zhao2019, yi2019sampling, ma2018esmm}, highlighting the value of shared representations while carefully managing interference. Our work follows this line but specifically targets the joint modeling of podcast ads and promotions within a single framework.

\textbf{Joint Optimization and Task Relatedness.}
A central challenge in MTL is trading off objectives that may conflict. Viewing MTL as multi-objective optimization provides principled ways to navigate Pareto trade-offs \cite{Sener2018,Lin2019, carmel2020moro, zheng2022multiobjective, jannach2023survey}. Another line of work studies when tasks should be learned together, showing that task affinity or relatedness strongly affects transfer \cite{Standley2020}. In our setting, we unify advertising and promotions within one model because they share user and content signals, while still needing to control cross-objective interference.

\textbf{Mitigating Negative Transfer.}
Negative transfer arises when gradients from different objectives conflict. Industry-ready approaches include learning to weight task losses (e.g., uncertainty weighting) \cite{Kendall2018}, gradient balancing/normalization \cite{chen2018gradnorm}, and gradient surgery to resolve conflicts (PCGrad) \cite{yu2020gradient}, as well as work on stabilizing large-scale multitask ranking models in production \cite{tang2023stability}. Architectural remedies such as MMoE share experts with task-specific gating to reduce interference at scale \cite{Ma2018}, and PLE introduces progressive shared/specific towers to further curb negative transfer in recommendation tasks \cite{tangPLE2020}. Our approach combines unified modeling with imbalance-aware training and careful sharing to retain positive transfer while limiting interference.

\section{System Evolution and Architecture}
We evolved from specialized models to a unified multi-task learning (MTL) framework that jointly optimizes podcast-related ad and promotion objectives. We first summarize the baselines and then formalize the joint ads--promotions model, including task definitions and training setup.

\subsection{Baseline Models and Initial Approaches}
Figure~2A shows our initial \emph{promotions-only multi-task model}. Each training example is an impression of a podcast promotion shown to a user. A shared feature encoder (with post-batch norm application) feeds task-specific towers---stacked MLPs---that predict user--podcast interactions (e.g., stream, click, like, follow) for promotions.

The encoder consumes four feature groups: (1) \emph{user} signals (historical listening, follows, search interactions, high-level profile attributes), (2) \emph{content} signals (show and episode identifiers, learned embeddings, genres, topics), (3) \emph{context} (time, surface, session state), and (4) \emph{promotion} metadata (slot, layout, campaign). We also considered Mixture-of-Experts (MoE) ~\cite{Ma2018, lin2024, mergingMTL, shazeer2017} variants, but the shared-bottom model served as the main production baseline.

Two intermediate approaches are shown in Figures~2A and 2B:
\begin{enumerate}
  \item \textbf{Promotions model for ad cold-start.} We reused the promotions model to score ad impressions. This enabled rapid launches for new ad objectives but ignored ad-specific features (e.g., creative type, campaign) and user--ad interaction patterns.
  \item \textbf{Single-task ads model.} We built an \emph{ads-only} model trained across all podcast ad surfaces and creatives (audio, video, display). It used similar user, content, and context features, plus ad-specific metadata (creative ID, format, campaign, slot). Despite rich ad logs, this single-task approach struggled to balance diverse business objectives effectively and support future goals requiring
  learning from all on-platform podcast interactions.
\end{enumerate}

Maintaining separate data pipelines and models increased engineering overhead and limited our ability to exploit shared structure across tasks, motivating a unified solution.

\subsection{Problem Formulation for Joint Ads--Promotions Modeling}
We treat targeting as predicting multiple per-impression outcomes for a user--podcast pair $(u, c)$ in context $x$ (e.g., surface, time, device). Let $\mathcal{T}$ be the set of binary prediction tasks, including:
\begin{itemize}
  \item \textit{PromotionStream}: Stream after a promotion impression;
  \item \textit{AdStream}: Stream after an ad impression;
  \item \textit{Click}: Click on a promotion or ad;
  \item \textit{Like} or \textit{Follow}: Like / follow of a promoted podcast.
\end{itemize}

For each task $t \in \mathcal{T}$, we observe a binary label $y_t \in \{0,1\}$. Given input features $x$, the model produces task-specific probabilities $p_t(x) = f_{\theta,t}(x)$, with shared and task-specific parameters $\theta$.

The unified model (Figure~2C) consists of:
\begin{itemize}
  \item a shared encoder $h_{\phi}(x)$ that maps user, content, context, and creative features into a joint representation $z = h_{\phi}(x)$;
  \item task-specific towers $g_{\psi_t}(z)$ that map $z$ to logits for each task $t$.
\end{itemize}
The predicted probability for task $t$ is
\[
p_t(x) = \sigma\bigl(g_{\psi_t}(h_{\phi}(x))\bigr),
\]
where $\sigma(\cdot)$ is the sigmoid function. Architecturally, the shared encoder mirrors the promotions baseline but incorporates ads-specific features and includes both ads and promotions tasks in $\mathcal{T}$, enabling joint learning over all podcast-related interactions while retaining task-specific capacity.

\subsection{Optimization and Loss Balancing}
We optimize binary cross-entropy losses over all tasks in $\mathcal{T}$, but with two design choices to control transfer between channels: (1) \emph{adaptive loss masking} from ads to promotions, and (2) \emph{source-balanced sampling} between promotions and ads.

Let $\mathcal{T}^{\mathrm{P}}$ and $\mathcal{T}^{\mathrm{A}}$ denote the sets of promotion and ad tasks respectively, with $\mathcal{T} = \mathcal{T}^{\mathrm{P}} \cup \mathcal{T}^{\mathrm{A}}$. We write $\mathcal{D}^{\mathrm{P}}$ and $\mathcal{D}^{\mathrm{A}}$ for the corresponding sets of promotion and ad impressions, and $\mathcal{D} = \mathcal{D}^{\mathrm{P}} \cup \mathcal{D}^{\mathrm{A}}$. Each impression $x \in \mathcal{D}$ has a source label $s(x) \in \{\mathrm{P}, \mathrm{A}\}$.

We define a binary mask $m_{s,t}$ that dictates whether task $t$ should incur loss on an impression from source $s$:
\[
m_{s,t} =
\begin{cases}
0, & \text{if } s = \mathrm{A} \text{ and } t \in \mathcal{T}^{\mathrm{P}}, \\
1, & \text{otherwise.}
\end{cases}
\]
This implements \emph{directional transfer}: promotion impressions update both promotion and ad towers, while ad impressions update only ad towers. The overall training objective is
\[
\mathcal{L}
= \sum_{t \in \mathcal{T}} \lambda_t \,
   \mathbb{E}_{(x, y_t) \sim \mathcal{D}}
   \Bigl[ m_{s(x),t} \, \ell_{\mathrm{BCE}}\bigl(y_t, p_t(x)\bigr) \Bigr],
\]
where $\lambda_t$ is a non-negative weight for task $t$ (set to 1 in our deployment) and $\ell_{\mathrm{BCE}}$ is the binary cross-entropy loss. In practice, the mask prevents ad-specific signals from directly shaping promotion towers, while still allowing promotion signals to aid ads, which is valuable given the relative data sparsity on some ad objectives.

To ensure parity between channels, we use \emph{source-balanced sampling}: each mini-batch is constructed so that roughly $50\%$ of impressions come from $\mathcal{D}^{\mathrm{P}}$ and $50\%$ from $\mathcal{D}^{\mathrm{A}}$. This keeps gradients from promotions and ads at comparable scales and avoids the joint model collapsing toward the higher-volume source.

\begin{figure*}[h]
\includegraphics[scale=0.31]{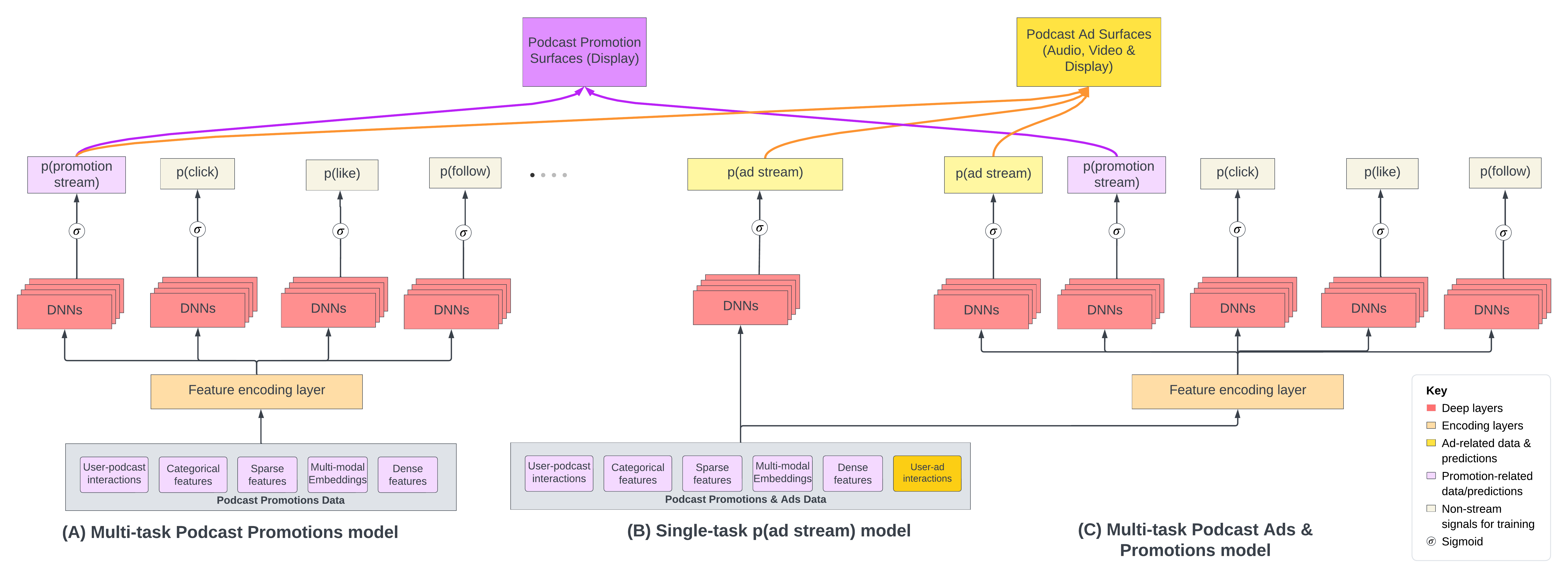}
\caption{(A) A promotions-only podcast model, used to serve ad stream predictions in the cold-start phase for the Ads objective. (B) Single-task pAdStream model incorporating both promotions and ads data. (C) Multi-task joint model for promotions and ads, serving both businesses.}\label{fig:camoe}
\end{figure*}

\section{Experiments and Results}
We compare the joint model with the promotions-only and ads-only baselines from Section~3.1. We outline the setup, then present offline and online results and summarize ablations.

\subsection{Experimental Setup}
\paragraph{Data and splits.}
We train on production logs from Spotify's podcast ads and promotions systems over a multi-month period. Impressions are temporally split into training, validation, and test sets: earlier days for training, intermediate days for validation, and the most recent days for testing. Ads and promotions impressions are pooled but retain channel labels and task-specific outcomes.

\paragraph{Evaluation metrics.}
Offline,  we use Average Precision (AP), which summarizes the precision--recall curve and is more informative than AUC-ROC under heavy class imbalance. Online, we focus on:
\begin{itemize}
  \item Effective Cost-Per-Stream (eCPS): ad spend divided by resulting podcast streams;
  \item Stream rate (i2s): impression-to-stream rate;
  \item Click-through rate (CTR).
\end{itemize}
Metrics are reported for all podcasts and for \emph{less-streamed creators} (shows with fewer than 5{,}000 streams), a segment strongly affected by cold-start.

\paragraph{Training details.}
All models share the same optimizer (Adam) and learning-rate schedule. Hyperparameters are tuned using validation AP on stream tasks.

\subsection{Offline Evaluation Results}
Table~\ref{tab:offline_ap} compares the multi-objective promo--ads model with the production baseline and alternative task groupings. The unified ``Promo + Ads 5-task MTL'' model provides the strongest performance.  

\begin{table}[t]
  \centering
  \caption{Average Precision (AP) comparison across configurations. Relative change to the baseline promotions model (Figure~2A).}
  \label{tab:offline_ap}
  \begin{tabular}{lcc}
    \toprule
    Task Setup                 & Promotions AP & Ads AP \\
    \midrule
    Promo Stream head-only     & $-7.9\%$      & $-8.8\%$ \\
    Ads Stream head-only       & $-65.2\%$     & $+27.0\%$ \\
    Ads Stream + ANC heads     & $-64.8\%$     & $+46.5\%$ \\
    Promo + Ads 5-task MTL     & $+4.5\%$      & $+50.2\%$ \\
    \bottomrule
  \end{tabular}
\end{table}

Relative to the promotions-only baseline, the joint model improves Promotions AP by $+4.5\%$ and Ads AP by $+50.2\%$. Ads-only configurations, even with ancillary heads, remain much weaker on promotions and still fall short of the joint model on ads, indicating that cross-channel transfer between promotions and ads is critical.

\subsection{Effect of Ancillary Heads}
The joint model includes ancillary heads for clicks, likes, and follows (ANC). Table~\ref{tab:offline_ap} shows that adding ANC heads to the ads-only model increases Ads AP from $+27\%$ to $+46.5\%$ relative to baseline, confirming that modeling intermediate engagement signals benefits stream prediction. However, this ads-only configuration severely degrades Promotions AP (around $-65\%$), indicating that ancillary heads alone are insufficient without promotions data.

In the unified MTL setting, ANC heads over both ads and promotions improve AP for \emph{both} channels. Ancillary labels are most useful when combined with cross-channel training, allowing the shared encoder to learn richer user and content representations.

\subsection{Online A/B Test Results}
We ran a budget-split A/B test across 180+ markets, comparing the 5-task joint model with the baseline that uses the promotions model for ad cold-start (Figure~2A).

\begin{table}[t]
  \centering
  \caption{A/B test results for all and less-streamed podcast creators (p-value $< 0.05$). Less-streamed podcasts have fewer than 5{,}000 streams. Relative change to the baseline (Figure~2A).}
  \label{tab:ab_results}
  \begin{tabular}{lcccc}
    \toprule
    Segment                & i2s   & eCPS  & CTR   & \# streams \\
    \midrule
    All podcasts           & $+18\%$ & $-20\%$ & $+10\%$ & $+18\%$ \\
    Less-streamed creators & $+24\%$ & $-22\%$ & $+9\%$  & $+27\%$ \\
    \bottomrule
  \end{tabular}
\end{table}

The joint model improves impression-to-stream rate, click-through rate, and cost-efficiency simultaneously. Gains are largest for less-streamed creators, with a $22\%$ eCPS reduction and $27\%$ more streams, proving this approach particularly effective for cold-start content.

\subsubsection{Cold-Start Performance}
To better understand how the joint model behaves across podcasts of different popularity levels, we further segment results by Spotify's \emph{stream tiers}. Podcasts are grouped into eight tiers based on the number of listening hours (longer than 60 seconds) accumulated over a rolling 30-day window. For our purposes, Tiers~0--2 correspond to \emph{high-stream} podcasts, while Tiers~3--5 capture \emph{low-stream} shows, aligned with less-streamed creator segment.

When we re-evaluate the A/B test by tiers, we observe markedly large improvements for lower-streamed podcasts. For high-stream tiers, the relative improvement in \textit{i2s} grows from approximately $+7\%$ (Tier~0) to $+20\%$ (Tier~2), while mean \textit{CPS} decreases by $4$--$17\%$. In contrast, low-stream tiers see substantially larger effects: \textit{i2s} improves by roughly $+27\%$ (Tier~3), $+33\%$ (Tier~4), and up to $+60\%$ for Tier~5, with corresponding CPS reductions of about $20\%$, $24\%$, and $38\%$, respectively.
This monotonic pattern---larger relative gains as we move from Tier~0 to Tier~5---provides strong evidence that the unified model is particularly effective in cold-start and low-stream regimes, where data is sparse and traditional siloed models struggle.

\section{Conclusion}

This paper presents the successful development and deployment of a unified multi-task model for podcast ad and promotion targeting at Spotify. Our joint optimization approach markedly improves upon traditional siloed models by effectively leveraging transfer learning; pre-training on extensive advertising data enables strong performance across diverse tasks, including promotions, particularly in cold-start scenarios.

Key lessons from this initiative highlight the power of unifying disparate yet related recommendation tasks, which not only unlocks significant performance gains but also fosters crucial organizational synergies, such as improved cross-team collaboration and strategic alignment by breaking down previously siloed efforts. Furthermore, leveraging transfer learning within such a joint model effectively mitigates cold-start issues for new content and objectives. The model's capacity to simultaneously enhance diverse business objectives—spanning ad streams, ad clicks, and promotional streams—with substantial gains suggests operation nearer to a Pareto optimal frontier \cite{Lin2019}. While our study focuses on podcasts, the approach naturally extends to other verticals (e.g., music, audiobooks, video) where ads and organic promotions share user and content representations.

\vfill
\eject

\bibliographystyle{ACM-Reference-Format}
\balance
\bibliography{mtl-bib}

@article{Caruana1997,
  author  = {Caruana, Rich},
  title   = {Multitask Learning},
  journal = {Machine Learning},
  year    = {1997},
  volume  = {28},
  pages   = {41--75}
}

@inproceedings{Collobert2008,
  author    = {Collobert, Ronan and Weston, Jason},
  title     = {A Unified Architecture for Natural Language Processing: Deep Neural Networks with Multitask Learning},
  year      = {2008},
  booktitle = {Proceedings of the 25th International Conference on Machine Learning},
  address   = {Helsinki, Finland},
  publisher = {ACM},
  pages     = {160--167},
  doi       = {10.1145/1390156.1390177},
  url       = {https://doi.org/10.1145/1390156.1390177}
}

@inproceedings{Covington2016,
  author    = {Covington, Paul and Adams, Jay and Sargin, Emre},
  title     = {Deep Neural Networks for YouTube Recommendations},
  year      = {2016},
  booktitle = {Proceedings of the 10th ACM Conference on Recommender Systems},
  address   = {Boston, MA, USA},
  publisher = {IEEE},
  doi       = {10.1145/2959100.2959190},
  url       = {https://doi.org/10.1145/2959100.2959190}
}

@inproceedings{Kendall2018,
  author    = {Kendall, Alex and Gal, Yarin and Cipolla, Roberto},
  title     = {Multi-task Learning Using Uncertainty to Weigh Losses for Scene Geometry and Semantics},
  year      = {2018},
  booktitle = {2018 IEEE/CVF Conference on Computer Vision and Pattern Recognition (CVPR)},
  address   = {Salt Lake City, UT, USA},
  publisher = {IEEE},
  doi       = {10.1109/CVPR.2018.00781},
  url       = {https://doi.org/10.1109/CVPR.2018.00781}
}

@inproceedings{Lin2019,
  author    = {Lin, Xi and Zhen, Hui-Ling and Li, Zhenhua and Zhang, Qingfu and Kwong, Sam},
  title     = {Pareto Multi-Task Learning},
  year      = {2019},
  booktitle = {Proceedings of the 33rd International Conference on Neural Information Processing Systems},
  address   = {Vancouver, Canada},
  publisher = {ACM},
  pages     = {12060--12070},
  doi       = {10.5555/3454287.3455367},
  url       = {https://doi.org/10.5555/3454287.3455367}
}

@inproceedings{Long2017,
  author    = {Long, Mingsheng and Cao, Zhangjie and Wang, Jianmin and Yu, Philip S.},
  title     = {Learning Multiple Tasks with Multilinear Relationship Networks},
  year      = {2017},
  booktitle = {Proceedings of the 31st International Conference on Neural Information Processing Systems},
  address   = {Long Beach, CA, USA},
  publisher = {ACM},
  pages     = {1593--1602},
  doi       = {10.5555/3294771.3294923},
  url       = {https://doi.org/10.5555/3294771.3294923}
}

@inproceedings{Ma2018,
  author    = {Ma, Jiaqi and Zhao, Zhe and Yi, Xinyang and Chen, Jilin and Hong, Lichan and Chi, Ed},
  title     = {Modeling Task Relationships in Multi-Task Learning with Multi-Gate Mixture-of-Experts},
  year      = {2018},
  booktitle = {Proceedings of the 24th ACM SIGKDD International Conference on Knowledge Discovery and Data Mining},
  address   = {London, United Kingdom},
  publisher = {ACM},
  pages     = {1930--1939},
  doi       = {10.1145/3219819.3220007},
  url       = {https://doi.org/10.1145/3219819.3220007}
}

@inproceedings{verma2025,
author = {Verma, Shivam and Chen, Vivian and Mei, Darren},
title = {An Audio-centric Multi-task Learning Framework for Streaming Ads Targeting on Spotify},
year = {2025},
isbn = {9798400714542},
publisher = {Association for Computing Machinery},
address = {New York, NY, USA},
url = {https://doi.org/10.1145/3711896.3737190},
doi = {10.1145/3711896.3737190},
booktitle = {Proceedings of the 31st ACM SIGKDD Conference on Knowledge Discovery and Data Mining V.2},
pages = {4945--4955},
numpages = {11},
location = {Toronto ON, Canada},
series = {KDD '25}
}

@inproceedings{tangPLE2020,
author = {Tang, Hongyan and Liu, Junning and Zhao, Ming and Gong, Xudong},
title = {Progressive Layered Extraction (PLE): A Novel Multi-Task Learning (MTL) Model for Personalized Recommendations},
year = {2020},
isbn = {9781450375832},
publisher = {Association for Computing Machinery},
address = {New York, NY, USA},
url = {https://doi.org/10.1145/3383313.3412236},
doi = {10.1145/3383313.3412236},
booktitle = {Proceedings of the 14th ACM Conference on Recommender Systems},
pages = {269--278},
numpages = {10},
location = {Virtual Event, Brazil},
series = {RecSys '20}
}

@article{yu2020gradient,
  title={Gradient Surgery for Multi-Task Learning},
  author={Yu, Tianhe and Kumar, Saurabh and Gupta, Abhishek and Levine, Sergey and Hausman, Karol and Finn, Chelsea},
  journal={Advances in Neural Information Processing Systems},
  volume={33},
  pages={5824--5836},
  year={2020}
}

@inproceedings{chen2018gradnorm,
  title={GradNorm: Gradient Normalization for Adaptive Loss Balancing in Deep Multitask Networks},
  author={Chen, Zhao and Badrinarayanan, Vijay and Lee, Chen-Yu and Rabinovich, Andrew},
  booktitle={International Conference on Machine Learning},
  pages={794--803},
  year={2018},
  organization={PMLR}
}

@inproceedings{Sener2018,
  author    = {Sener, Ozan and Koltun, Vladlen},
  title     = {Multi-Task Learning as Multi-Objective Optimization},
  year      = {2018},
  booktitle = {Proceedings of the 32nd International Conference on Neural Information Processing Systems},
  address   = {Montr{\'e}al, Canada},
  publisher = {ACM},
  pages     = {525--536},
  doi       = {10.5555/3326943.3326992},
  url       = {https://doi.org/10.5555/3326943.3326992}
}

@inproceedings{Standley2020,
  author    = {Standley, Trevor and Zamir, Amir R. and Chen, Dawn and Guibas, Leonidas and Malik, Jitendra and Savarese, Silvio},
  title     = {Which Tasks Should Be Learned Together in Multi-Task Learning?},
  year      = {2020},
  booktitle = {Proceedings of the 37th International Conference on Machine Learning},
  series    = {ICML '20}
}

@inproceedings{googleads2022,
author = {Ma, Ning and Ispir, Mustafa and Li, Yuan and Yang, Yongpeng and Chen, Zhe and Cheng, Derek Zhiyuan and Nie, Lan and Barman, Kishor},
title = {An Online Multi-task Learning Framework for Google Feed Ads Auction Models},
year = {2022},
isbn = {9781450393850},
publisher = {Association for Computing Machinery},
address = {New York, NY, USA},
url = {https://doi.org/10.1145/3534678.3539055},
doi = {10.1145/3534678.3539055},
booktitle = {Proceedings of the 28th ACM SIGKDD Conference on Knowledge Discovery and Data Mining},
pages = {3477--3485},
numpages = {9},
location = {Washington DC, USA},
series = {KDD '22}
}

@misc{lin2024,
      title={MoMa: Efficient Early-Fusion Pre-training with Mixture of Modality-Aware Experts}, 
      author={Xi Victoria Lin and Akshat Shrivastava and Liang Luo and Srinivasan Iyer and Mike Lewis and Gargi Ghosh and Luke Zettlemoyer and Armen Aghajanyan},
      year={2024},
      eprint={2407.21770},
      archivePrefix={arXiv},
      primaryClass={cs.AI},
      url={https://arxiv.org/abs/2407.21770}, 
}

@inproceedings{mergingMTL,
  title={Merging Multi-Task Models via Weight-Ensembling Mixture of Experts},
  author={Tang, Anke and Shen, Li and Luo, Yong and Yin, Nan and Zhang, Lefei and Tao, Dacheng},
  year={2024},
  booktitle={Forty-first International Conference on Machine Learning}
}

@inproceedings{Zhao2019,
  author    = {Zhao, Zhe and Hong, Lichan and Wei, Li and Chen, Jilin and Nath, Aniruddh and Andrews, Shawn and Kumthekar, Aditee and Sathiamoorthy, Maheswaran and Yi, Xinyang and Chi, Ed},
  title     = {Recommending What Video to Watch Next: A Multitask Ranking System},
  year      = {2019},
  booktitle = {Proceedings of the 13th ACM Conference on Recommender Systems},
  address   = {Copenhagen, Denmark},
  publisher = {ACM},
  pages     = {43--51},
  doi       = {10.1145/3298689.3346997},
  url       = {https://doi.org/10.1145/3298689.3346997}
}

@inproceedings{shazeer2017,
title={ Outrageously Large Neural Networks: The Sparsely-Gated Mixture-of-Experts Layer},
author={Noam Shazeer and *Azalia Mirhoseini and *Krzysztof Maziarz and Andy Davis and Quoc Le and Geoffrey Hinton and Jeff Dean},
booktitle={International Conference on Learning Representations},
year={2017},
url={https://openreview.net/forum?id=B1ckMDqlg}
}

@inproceedings{lirank2024,
  title={LiRank: Industrial Large Scale Ranking Models at LinkedIn},
  author={Borisyuk, Fedor and Zhou, Mingzhou and Song, Qingquan and Zhu, Siyu and Tiwana, Birjodh and Parameswaran, Ganesh and Dangi, Siddharth and Hertel, Lars and Xiao, Qiang Charles and Hou, Xiaochen and others},
  booktitle={Proceedings of the 30th ACM SIGKDD Conference on Knowledge Discovery and Data Mining},
  pages={4804--4815},
  year={2024}
}

@article{moo1,
author = {Wu, Haolun and Ma, Chen and Mitra, Bhaskar and Diaz, Fernando and Liu, Xue},
title = {A Multi-Objective Optimization Framework for Multi-Stakeholder Fairness-Aware Recommendation},
year = {2022},
issue_date = {April 2023},
publisher = {Association for Computing Machinery},
address = {New York, NY, USA},
volume = {41},
number = {2},
issn = {1046-8188},
url = {https://doi.org/10.1145/3564285},
doi = {10.1145/3564285},
journal = {ACM Trans. Inf. Syst.},
month = dec,
articleno = {47},
numpages = {29}
}

@inproceedings{moo2,
author = {Jeunen, Olivier and Mandav, Jatin and Potapov, Ivan and Agarwal, Nakul and Vaid, Sourabh and Shi, Wenzhe and Ustimenko, Aleksei},
title = {Multi-Objective Recommendation via Multivariate Policy Learning},
year = {2024},
isbn = {9798400705052},
publisher = {Association for Computing Machinery},
address = {New York, NY, USA},
url = {https://doi.org/10.1145/3640457.3688132},
doi = {10.1145/3640457.3688132},
booktitle = {Proceedings of the 18th ACM Conference on Recommender Systems},
pages = {712--721},
numpages = {10},
location = {Bari, Italy},
series = {RecSys '24}
}

@inproceedings{airbnb2023,
author = {Tan, Chun How and Chan, Austin and Haldar, Malay and Tang, Jie and Liu, Xin and Abdool, Mustafa and Gao, Huiji and He, Liwei and Katariya, Sanjeev},
title = {Optimizing Airbnb Search Journey with Multi-task Learning},
year = {2023},
isbn = {9798400701030},
publisher = {Association for Computing Machinery},
address = {New York, NY, USA},
url = {https://doi.org/10.1145/3580305.3599881},
doi = {10.1145/3580305.3599881},
booktitle = {Proceedings of the 29th ACM SIGKDD Conference on Knowledge Discovery and Data Mining},
pages = {4872--4881},
numpages = {10},
location = {Long Beach, CA, USA},
series = {KDD '23}
}

@inproceedings{cheng2016wide,
  author    = {Heng{-}Tze Cheng and Levent Koc and Jeremiah Harmsen and Tal Shaked
               and Tushar Chandra and Hrishi Aradhye and Glen Anderson and
               Greg Corrado and Wei Chai and Mustafa Ispir and Rohan Anil and
               Zakaria Haque and Lichan Hong and Vihan Jain and Xiaobing Liu and
               Hemal Shah},
  title     = {Wide \& Deep Learning for Recommender Systems},
  booktitle = {Proceedings of the 1st Workshop on Deep Learning for Recommender Systems (DLRS)},
  pages     = {7--10},
  year      = {2016},
  publisher = {ACM}
}

@inproceedings{ma2018esmm,
  author    = {Xiao Ma and Liqin Zhao and Guan Huang and Zhi Wang and Zelin Hu and
               Xiaoqiang Zhu and Kun Gai},
  title     = {Entire Space Multi-Task Model: An Effective Approach for Estimating Post-Click Conversion Rate},
  booktitle = {Proceedings of the 41st International ACM SIGIR Conference on Research and Development in Information Retrieval},
  pages     = {1137--1140},
  year      = {2018},
  publisher = {ACM}
}

@inproceedings{yi2019sampling,
  author    = {Xinyang Yi and Ji Yang and Lichan Hong and Derek Zhiyuan Cheng and
               Lukasz Heldt and Aditee Kumthekar and Zhe Zhao and Li Wei and
               Ed H. Chi},
  title     = {Sampling-Bias-Corrected Neural Modeling for Large Corpus Item Recommendations},
  booktitle = {Proceedings of the 13th ACM Conference on Recommender Systems},
  pages     = {269--277},
  year      = {2019},
  publisher = {ACM}
}

@article{zheng2022multiobjective,
  author    = {Yong Zheng and David Xuejun Wang},
  title     = {A Survey of Recommender Systems with Multi-Objective Optimization},
  journal   = {Neurocomputing},
  volume    = {474},
  pages     = {141--153},
  year      = {2022},
  publisher = {Elsevier}
}

@article{jannach2023survey,
  author    = {Dietmar Jannach and Himan Abdollahpouri},
  title     = {A Survey on Multi-Objective Recommender Systems},
  journal   = {Frontiers in Big Data},
  volume    = {6},
  pages     = {1157899},
  year      = {2023}
}

@inproceedings{carmel2020moro,
  author    = {David Carmel and Elad Haramaty and Arnon Lazerson and Liane Lewin{-}Eytan},
  title     = {Multi-Objective Ranking Optimization for Product Search Using Stochastic Label Aggregation},
  booktitle = {Proceedings of The Web Conference 2020},
  pages     = {373--383},
  year      = {2020},
  publisher = {ACM}
}

@inproceedings{tang2023stability,
  author    = {Jiaxi Tang and Yoel Drori and Daryl Chang and
               Maheswaran Sathiamoorthy and Justin Gilmer and Li Wei and
               Xinyang Yi and Lichan Hong and Ed H. Chi},
  title     = {Improving Training Stability for Multitask Ranking Models in Recommender Systems},
  booktitle = {Proceedings of the 29th {ACM} {SIGKDD} Conference on Knowledge Discovery and Data Mining},
  pages     = {4882--4893},
  year      = {2023},
  publisher = {ACM}
}



\end{document}